# Simulation Study of an Energy-Efficient Time Synchronization Scheme based on Source Clock Frequency Recovery in Asymmetric Wireless Sensor Networks

Kyeong Soo Kim, Sanghyuk Lee, and Eng Gee Lim

Dept. of Electrical and Electronic Eng., Xi'an Jiaotong-Liverpool University, Suzhou 215123, P. R. China

**ABSTRACT**

In this paper we report preliminary results of a simulation study on an energy-efficient time synchronization scheme based on source clock frequency recovery (SCFR) at sensor nodes in asymmetric wireless sensor networks (WSNs), where a head node — serving as a gateway between wired and wireless networks — is equipped with a powerful processor and supplied power from outlet, and sensor nodes — connected only through wireless channels — are limited in processing and battery-powered. In the SCFR-based WSN time synchronization scheme, we concentrate on battery-powered sensor nodes and reduce their energy consumption by minimizing the number of message transmissions from sensor nodes to the head node. Through simulation experiments we analyze the performance of the SCFR-based WSN time synchronization scheme, including the impact of SCFR on time synchronization based on two-way message exchange, and demonstrate the feasibility of the proposed time synchronization scheme.

## 1. INTRODUCTION

Time synchronization is one of critical components in wireless sensor network (WSN) operation, as it provides a common time frame among different nodes. It supports functions such as fusing data from different sensor nodes, time-based channel sharing and media access control (MAC) protocols, and coordinated sleep wake-up node scheduling mechanisms [1]. Sensor nodes are low-complexity, battery-powered devices, so energy efficiency is the key in designing schemes and protocols for WSNs.

In a typical WSN, a master/head node is equipped with a powerful processor, connected to both wired and wireless networks, and supplied power from outlet because it serves as a gateway between the WSN and a backbone and a center for fusion of sensory data from sensor nodes, which are limited in processing and electrical power because they are connected only through wireless channels and battery-powered. It is this asymmetry that we focus our study on; unlike existing schemes which save the power of all WSN nodes including the head (e.g., [2] and [3]), we concentrate on battery-powered sensor nodes — which are many in numbers — in minimizing energy consumption for time synchronization. Specifically, in this paper we discuss a time synchronization scheme based on the source clock frequency recovery (SCFR) [4], which was introduced in [5]; we minimize the number of message transmissions at sensor nodes because the energy for packet transmission is typically higher than that for packet reception [6].

## 2. REVIEW OF SCFR-BASED WSN TIME SYNCHRONIZATION [5]

Here we briefly review the proposed WSN time synchronization scheme and discuss related issues. The major idea is to allow unsynchronized slave clocks at sensor nodes but running at the same frequency as the master clock at a head node through the asynchronous SCFR schemes described in [4], while carrying out the two-way message exchange, which is unavoidable for the recovery of clock offset in the existence of propagation delay [7], using normal data packets to reduce the number of transmissions at sensor nodes. In this way, the head node can estimate time offsets of sensor nodes and correctly interpret the occurrence of data measurements with respect to its own master clock.

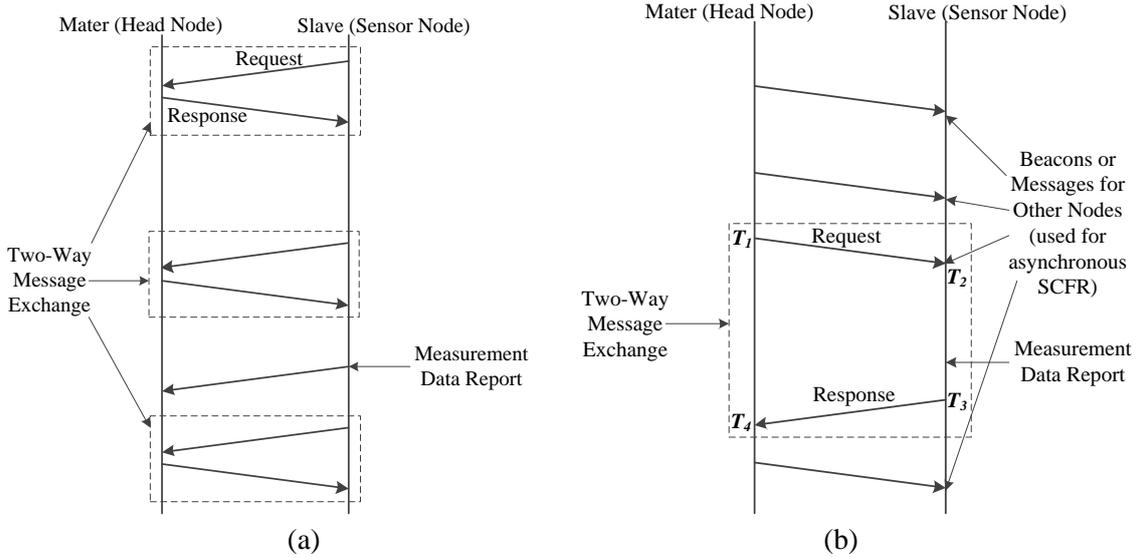

Fig. 1 Reducing message transmissions at sensor nodes: (a) A scheme based on two-way message exchange as in time-sync protocol for sensor networks (TPSN) [8] and (b) the proposed scheme.

Fig. 1 illustrates this idea in comparison to an ordinary two-way message exchange scheme shown in Fig. 1 (a): First, the proposed scheme shown in Fig. 1 (b) does not have periodic, dedicated two-way message exchange sessions with special control messages like "Request" and "Response"; instead, the two-way message exchange is carried out using an ordinary message (e.g., for measurement data) from a sensor node and the most recent message from the head, both of which have embedded timestamps. Secondly, the direction of two-way message exchange in the proposed scheme is reversed, i.e., it is the master that requests, not the slave, unlike the existing schemes; as a result, the master knows the current status of the slave clock, but the slave does not. So the information of slave clocks (i.e., time offsets with respect to the master clock) is centrally managed at the head node.

For operations like coordinated sleep wake-up node scheduling, the head node first adjusts the time for future operation (with respect to its own master clock) based on the time offset of a recipient sensor node before sending it to that node in the proposed scheme. In this way, even though sensor nodes in the network have clocks with different time offsets, their operations can be coordinated based on the common master clock in the head node.

Note that for the proposed scheme, the use of SCFR at a sensor node is critical in carrying out the two-way message exchange procedure: Unlike the ordinary procedure shown in Fig. 1 (a), the time period between the "Request" message and its corresponding "Response" message (i.e., $T_4$-$T_2$) can be quite long when the sensor node has no immediate measurement data to report back to the head as shown in Fig. 1 (b).

When the two clocks at the sensor and the head nodes run at the same frequency, we can easily calculate the clock offset $\theta$ and propagation delay $D$ as follows [8]:

$$\theta = \frac{(T_2-T_1)-(T_4-T_3)}{2} \quad (1)$$

$$D = \frac{(T_2-T_1)+(T_4-T_3)}{2} \quad (2)$$

where $T_1$ and $T_4$ represent the times measured by the head node clock, while $T_2$ and $T_3$ represent the times measured by the sensor node clock.

If the sensor clock has clock skew $R$ (i.e., the frequency ratio of the sensor clock to the head clock), however, Eqs. (1) and (2) are no longer exact estimates of $\theta$ and $D$. Therefore, the proposed time synchronization scheme demands a high-quality SCFR technique, especially when the frequency of data measurement/report is low.

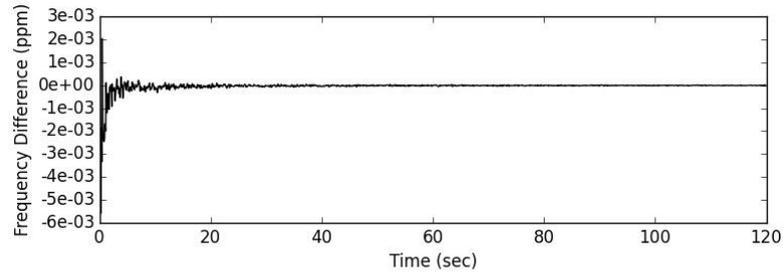
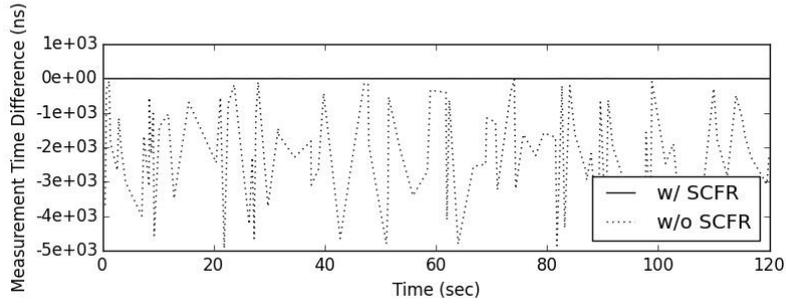

(a)

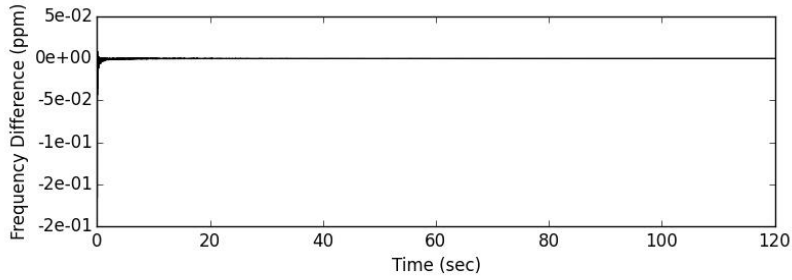
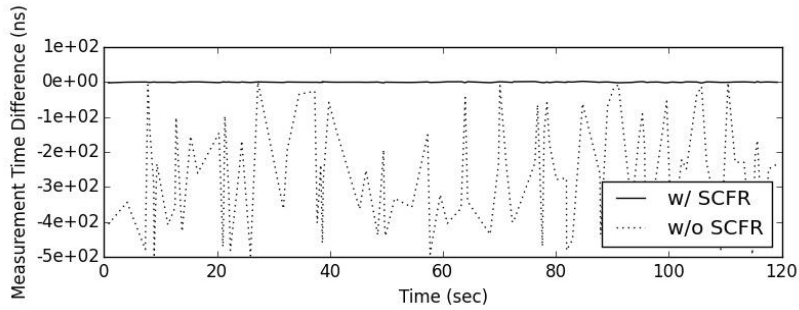

(b)

Fig. 2 Performance of SCFR at a sensor node and measurement time estimation at a head node for beacon interval of (a) 100 ms and (b) 10 ms.

## 3. SIMULATION RESULTS

Here we consider a simple WSN with one head node and one sensor node that are deployed 100 m from each other. We set the clock skew $R$ and the clock offset $\theta$ to $1 + 100$ ppm and 1 s, respectively. We model timestamp generation and reception noise with an i.i.d. Gaussian process having a standard deviation of 1 ns as in [9]. During the observation interval of 120 s, total 100 measurements are made where their corresponding data arrivals are modeled as a Poisson process. Also, to investigate the effect of the time period between the "Request" message and its corresponding "Response" message, we run simulations for two different values of beacon interval at the head node, i.e., 100 ms and 10 ms. For SCFR, we use the simple cumulative ratio scheme proposed in [4].

The results of SCFR at the sensor node and measurement time estimation at the head node (i.e., the estimate of $T_3$ in terms of the head node clock) are shown in Figs. 2 and 3. As discussed in Sec. 3, the results show that, without SCFR, the performance of measurement time estimation highly depends on beacon interval. The use of SCFR, on the other hand, removes this dependency, and the resulting measurement time difference is of the order of the noise standard deviation (i.e., ≈ 1 ns).

**4. SUMMARY**

In this paper we have provided preliminary simulation results in order to show the feasibility of the proposed energy-efficient time synchronization scheme for asymmetric WSNs. The results also demonstrate the importance of the use of SCFR in two-way message exchange procedure for measurement time estimation.